\documentclass{PoS}

\usepackage{amssymb}
\usepackage{amsmath}
\usepackage{amsthm}
\usepackage{makecell}

\title{Overview of top quark physics at the ep colliders}

\ShortTitle{Overview of top quark physics at the ep colliders}

\author{\speaker{Hao Sun}
        \thanks{for the LHeC working group}
        \\
        Institute of Theoretical Physics, School of Physics,\\
        Dalian University of Technology, \\
        No.2 Linggong Road, Dalian, Liaoning, 116024, P.R.China\\
        E-mail: \email{haosun@dlut.edu.cn}}

\abstract{
In this talk we present a short overview of top physics at the electron-proton (ep) colliders.
Currently, the proposed ep collider is the Large Hadron Electron Collider (LHeC),
which is a combination of 60 GeV electron beam and 7 TeV proton beam of the LHC tunnel.
This may be extended to Future Circular electron-hadron Collider (FCC-eh),
which features a 60 GeV (or higher) electron beam
with the 50 TeV proton beam from the Future Circular hadron Collider (FCC-hh).
Selected topics include but not limited to top structure function, top parton distribution functions,
top spin polarization, top electric charge, measurement of $\rm V_{tb}$,
anomalous $\rm tt\gamma$, ttZ, tbW, $\rm tq\gamma$, tqH couplings and CP phase of ttH coupling.
}

\FullConference{XXV International Workshop on Deep-Inelastic Scattering and Related Subjects\\
		3-7 April 2017\\
		University of Birmingham, UK}

\begin{document}

\section{Introduction}

Electron-proton (ep) colliders are hybrids between electron-position ($\rm e^+e^-$)
and proton-proton (pp) colliders, which consist of a hadron beam with an electron beam.
They provide a cleaner environment compared to the pp colliders
and higher center-of-mass (CMS) energies to the $\rm e^+e^-$ ones.
Currently, the proposed ep collider is the Large Hadron Electron Collider (LHeC),
which is a combination of 60 GeV electron beam and 7 TeV proton beam of the Large Hadron Collider (LHC).
It can deliver up to 100 $\rm fb^{-1}$ integrated luminosity per year at a CMS energy of around 1 TeV
and 1 $\rm ab^{-1}$ over its lifetime.
This may later be extended to Future Circular electron-hadron Collider (FCC-eh),
which features a 60 GeV (or higher) electron beam
with the 50 TeV proton beam from the Future Circular hadron Collider (FCC-hh).
This would result in a CMS energy up to 3.5 TeV.
The ep collider is advantageous for several reasons.
For example, it provides a clean environment with suppressed backgrounds from strong interactions,
therefore free from issues like pile-ups, multiple interactions.
In addition, the backward and forward scattering
can be disentangled due to the asymmetric of the initial states.
Such a facility will be very useful in understanding proton
and gluon interactions at very low x and very high $\rm Q^2$,
thus providing a much needed complementary information to the physics program of the LHC.
They would enable new precision studies of QCD,
and the precision determination of parton distributions functions (pdfs)
in a largely extended kinematic region. It would also provide additional
and sometimes unique ways for studying top and electroweak physics,
as well as Higgs and physics beyond the Standard Model (SM).

This talk focuses on some of the selected topics
on top physics at the LHeC and FCC-eh. They are summarized in Section.2,
which are arranged through different top quark production mechanisms:
the neutral current top production, the charged current top production, and top and Higgs related productions.
Finally, we end with a short summary.

\section{Selected topics on top physics at the ep colliders}

\subsection{Neutral current top production}

\begin{figure}[hbtp]
\centering
\includegraphics[scale=0.4]{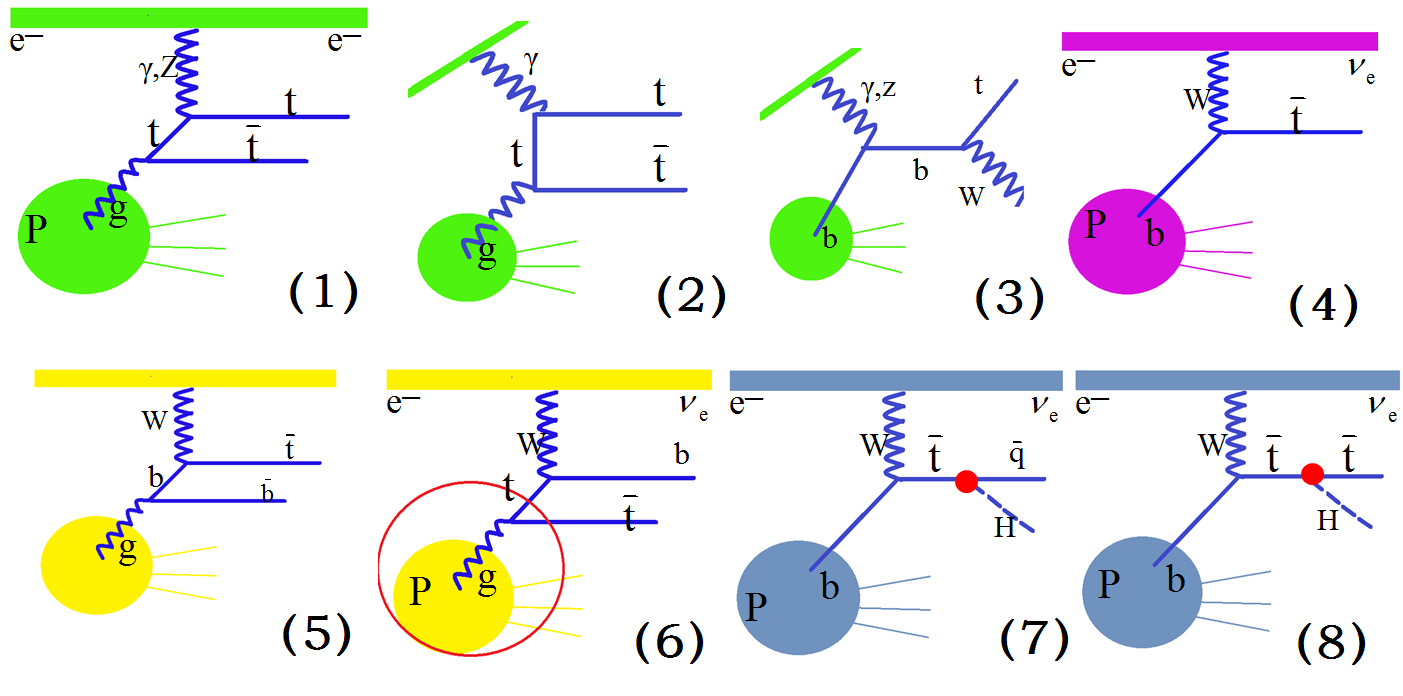}
\vspace{-0.3cm}
\caption{\label{Feynman}
Illustrated diagrams for NC, CC top productions and top and Higgs related productions.}
\end{figure}
We start from the neutral current (NC) top quark production.
For NC top production, there are mainly two modes.
One is the Deep Inelastic Scattering (DIS) and the other is photoproduction.
Single top and Top pair production occur in both modes.
Considering the 60 GeV electron beam and 7 TeV proton beam,
the DIS top pair cross section is about 23 fb (Fig.1(1)).
This is comparable to single top photoproduction which
is about 31 fb (Fig.1(3)). Both are smaller than the ep-based $\rm \gamma p$
collision where the top pair cross section is as large as 700 fb (Fig.1(2))
which means 70,000 top pairs can be obtained with a integrated luminosity of 100 $\rm fb^{-1}$.

DIS fermion pair production is sensitive to the gluon density in the proton.
By using DIS $\rm t\bar{t}$ production one can study the top component
of the structure function at especially small x region in the ep
project\cite{top_structure_Boroun}.
Top pair photoproduction can be used, for example, to measure the $\rm tt\gamma$ vertex, thus,
the electric charge of the top quark. Even though the production rate is lower than at the LHC,
the potential of measurement is better\cite{ttrttz_Bouzas}.
The highly energetic incoming photon couples only to the top quark,
so that the cross section depends directly on the $\rm tt\gamma$ vertex.
This is in contrast to the LHC where $\rm tt\gamma$ is probed through $\rm tt\gamma$ production,
where the outgoing photon could come from other charged sources such as the top decay products.
When the anomalous $\rm tt\gamma$ vertex is introduced, the top pair photoproduction
can also be used to set constraints on the Magnetic
and Electric Dipole moment of $\rm tt\gamma$ couplings\cite{ttrttz_Bouzas}.
Similarly but through DIS top pair production from Gluon-Z-boson fusion,
one can study the anomalous ttZ couplings\cite{ttrttz_Bouzas}, though with less sensitivity.
For the anomalous $\rm tq\gamma$ interaction, it is found that
ep-based $\rm \gamma p$ collider can provide a nice place to probe it\cite{tqr_Cakir}.
With 1 $\rm ab^{-1}$ integrated luminosity the $\rm tq\gamma$ coupling can be probed up to order of $\rm 10^{-3}$,
corresponding to Br($\rm t\to q\gamma$) of order $\rm 10^{-6}$, which shows quite nice feature\cite{LHeC_JPG2012}.

\subsection{Charged current top production}

Let's turn to charged current (CC) top quark production.
At the ep collider, single top quark is largely produced through CC mode.
Once the bottom quark pdf has been introduced, the top production
is triggered through this 2 to 2 scattering in Fig.1(4). This production can reach
1.73 pb at the LHeC and 15.3 pb at the FCC-eh, which are quite large.
Another source of single top production is the 2 to 3 process in Fig.1(5),
where the gluon splits into $\rm b\bar{b}$ pair,
while the $\rm\bar{b}$ quark is produced in the final state as a spectator.
This contribution can be included as a correction to the former 2 to 2 production.
One should be careful when combine them together,
an overlap terms should be subtracted\cite{top-spin-pol},
as the logarithmic part of the 2 to 3 process is already
covered the bottom quark pdf used to compute the 2 to 2 process.

The CC single top production can be measured
through W boson leptonic or hadronic decay modes.
The signal over background ratio can be larger than one for the hadronic mode
and even reach 11 through the leptonic decay mode.
The precise measurement of this production can be used to measure,
for example, the CKM matric element $\rm V_{tb}$. It is found that at the LHeC,
with luminosity of 100 $\rm fb^{-1}$, $\rm V_{tb}$
can be measured with a precision of $0.5\%$.
In addition, the top quark spin polarization can also be precisely measured
through this production\cite{top-spin-pol}:
$\rm 1/\Gamma_t d\Gamma/d\cos\theta = 1/2(1+A \alpha \cos\theta)$,
wher $\theta$ is the angle between the charged lepton (decay from the W boson)
and the spin quantisation axis in the top rest frame and A is the spin asymmetry.
The CC single top production is also a good way to measure the anomalous $\rm tbW$ couplings
at the ep project. A detailed study was performed in \cite{Wtb_Kumar},
in a model independent way, by means of the following effective CP conserving lagrangian:
\begin{equation}
\rm {\cal L}_{Wtb}=\frac{g}{\sqrt{2}}\left[W_\mu\bar{t}\gamma^\mu\left(V_{tb}f^L_1P_L+f^R_1P_R\right)b
-\frac{1}{2m_W}W_{\mu\nu}\bar{t}\sigma^{\mu\nu}\left(f^L_2P_L+f^R_2P_R\right)b\right] + h.c.
\end{equation}
where $\rm f^L_1(\equiv 1+\Delta f^L_1)$ and $\rm f^R_1$ are left- and right-handed vector couplings,
$\rm f^{L,R}_2$ are left- and right-handed tensor couplings,
$\rm W_{\mu\nu}=\partial_\mu W_\nu-\partial_\nu W_\mu$, $\rm P_{L,R}=\frac{1}{2}(1\mp\gamma_5)$
are left- and right-handed projection operators,
$\rm \sigma^{\mu\nu}=i/2(\gamma^\mu\gamma^\nu-\gamma^\nu\gamma^\mu)$ and $\rm g=2/\sin\theta_W$.
The fact that the left-handed vector current is one to a very good approximation,
and the vanishing of the other parameters corresponding to the SM case, as within the SM the
$\rm tbW$ vertex is purely left-handed.
Analyses were performed using a simulated event sample
corresponding to an integrated luminosity of 100 $\rm fb^{-1}$
for different systematic uncertainties.
Contours at 68\% and 95\% confidence level on two dimensional plane
for any coupling combination were presented in \cite{Wtb_Kumar}.
Results in comparison with others from Tevatron,
LHC and indirect one from $\rm B$ decays are shown in \cite{LHeC_pos}.
The analysis shows that the tbW vertex can be probed at the LHeC to a very high accuracy
and one can obtain comparable or better results from other determinations.
There is another CC top production mode present in Fig.1(6),
making it possible to consider quark density for the top quark,
since at very high scales the top may be considered "light".
A six-flavor scheme has been proposed in \cite{LHeC_JPG2012},
thus the LHeC offers new field of research for the top quark pdfs.

\subsection{Charged current top and Higgs related production}

Finally lets see the top and Higgs related productions at the ep colliders.
When the lagrangian
\begin{eqnarray}\label{lagrangian}
\rm {\cal L} = \kappa_{tuH} \bar{t}uH + \kappa_{tcH} \bar{t}cH + h.c.
\end{eqnarray}
is introduced, we can use this channel to study the FCNC tqH couplings.
The considered signal include top-Higgs FCNC couplings\cite{tqh_haosun} is
$\rm e^- p \to \nu_e \bar{t} \rightarrow \nu_e H \bar{q}$ (Fig.1(7)).
At 60 GeV and 7(50) TeV LHeC(FCC-eh),
the cross section is as large as 10.5(86.1) fb.
Another production include tqH coupling is $\rm e^- p\to \nu_e H b$. In this case,
the initial quark is a u(c)-quark instead of a b-quark. However, this cross section is
small compare to the former, which is only 0.81(3.8) fb, thus not considered here.
The decay channel we considered is H to $\rm b\bar{b}$,
$\rm e^- p \to \nu_e \bar{t} \rightarrow \nu_e H \bar{q} \to \nu_e b\bar{b} \bar{q}$,
therefore the signature is characterized by three (or more than three) jets
and a missing transverse momentum from the undetected neutrino.
Two of the jets should be tagged as b-jets. The combination of the two b-jets
should appear as a narrow resonance centered around the SM Higgs boson mass.
Together with the remaining light jet(s), they should be able to reconstruct a resonant top quark.
The main backgrounds come from both the reducible and irreducible ones.
The crucial irreducible backgrounds which yield exactly
the same final states to the signal are listed here, see,
$\rm e^-p \rightarrow \nu_e (H\rightarrow b \bar{b} ) j$, $\rm e^-p \rightarrow \nu_e (z\rightarrow b \bar{b} ) j$
which contain three QED couplings, $\rm e^-p \rightarrow \nu_e (g\rightarrow b \bar{b}) j$
which contains two QED couplings and two QCD couplings.
Notice here and bellow, j = g, u, $\rm \bar{u}$, d, $\rm \bar{d}$, c, $\rm \bar{c}$, s, $\rm \bar{s}$.
One source of the most important potentially reducible backgrounds are
$\rm e^-p \rightarrow \nu_e jjj$ and $\rm e^-p \rightarrow \nu_e jjb/\bar{b}$.
These are the irreducible and reducible backgrounds we considered.
\begin{figure}[hbtp]
\centering
\includegraphics[scale=0.23]{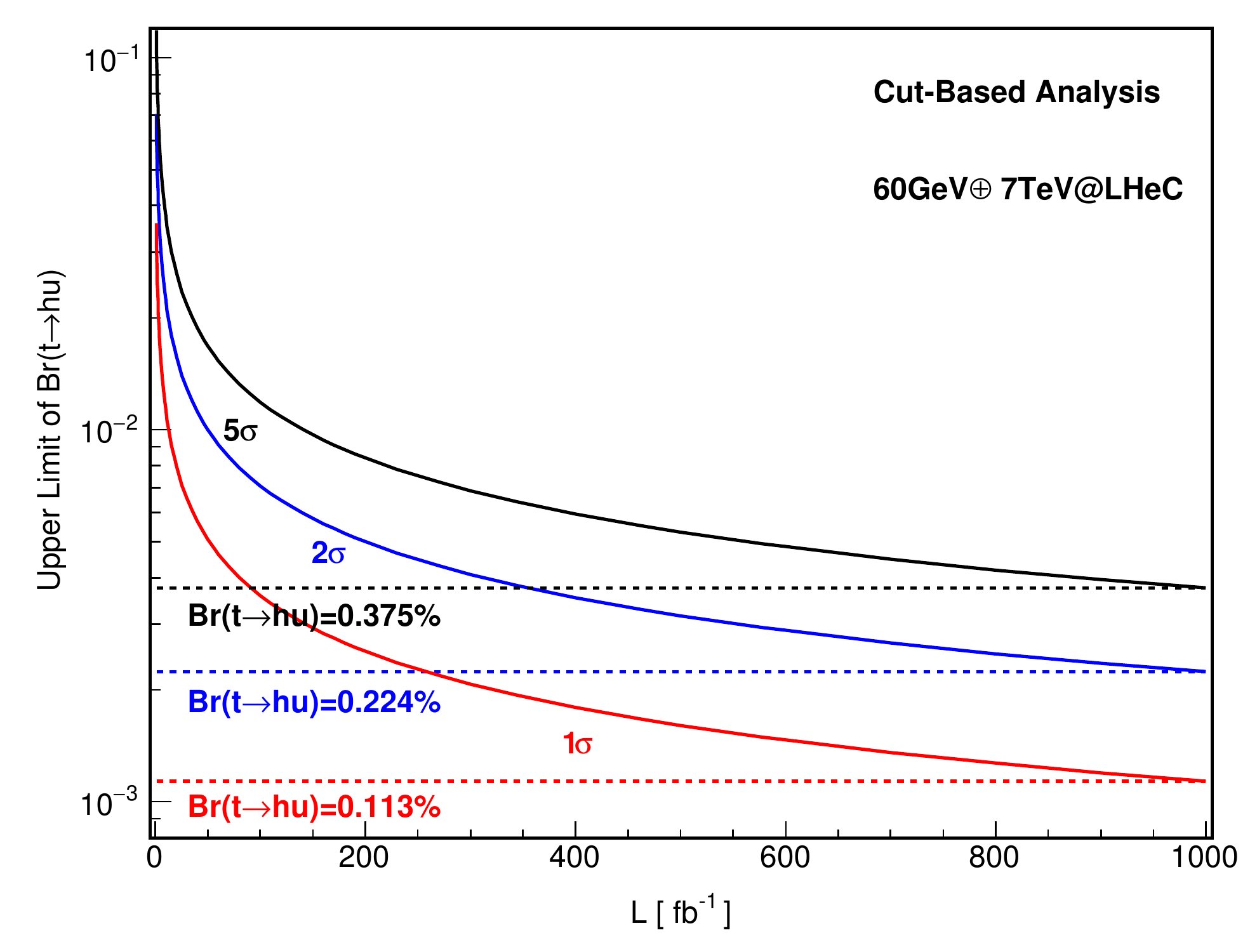}
\includegraphics[scale=0.23]{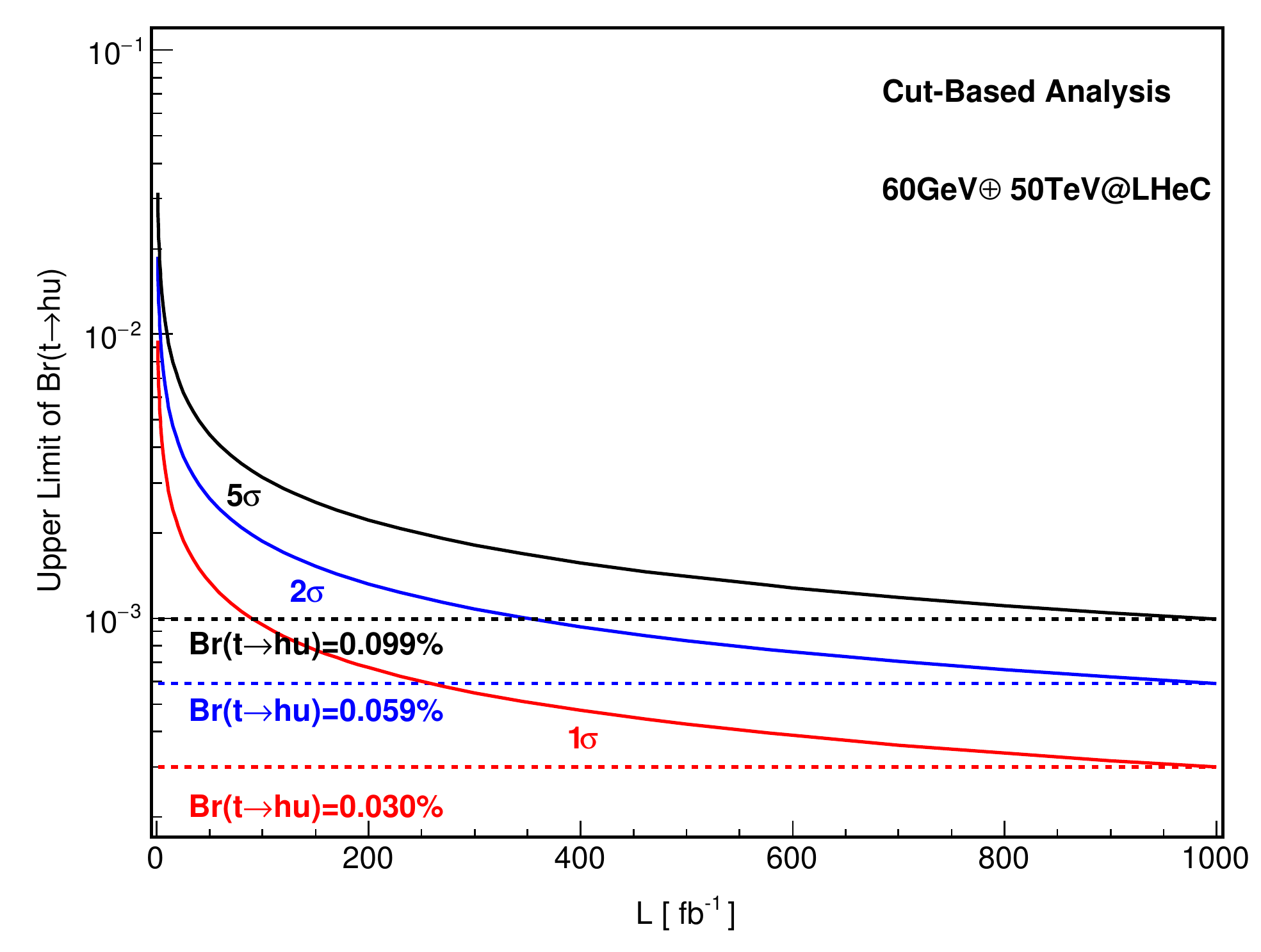}\\
\includegraphics[scale=0.23]{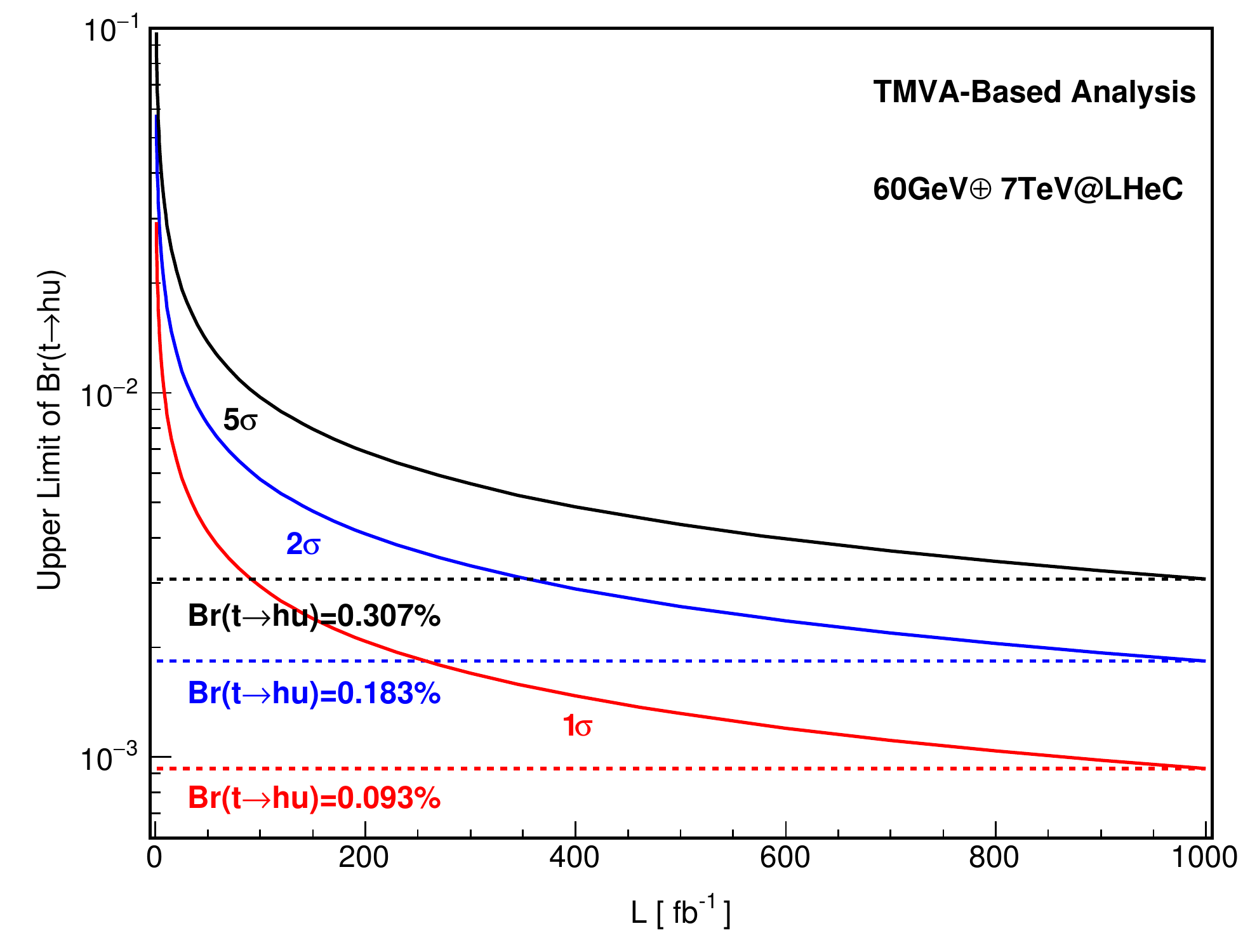}
\includegraphics[scale=0.23]{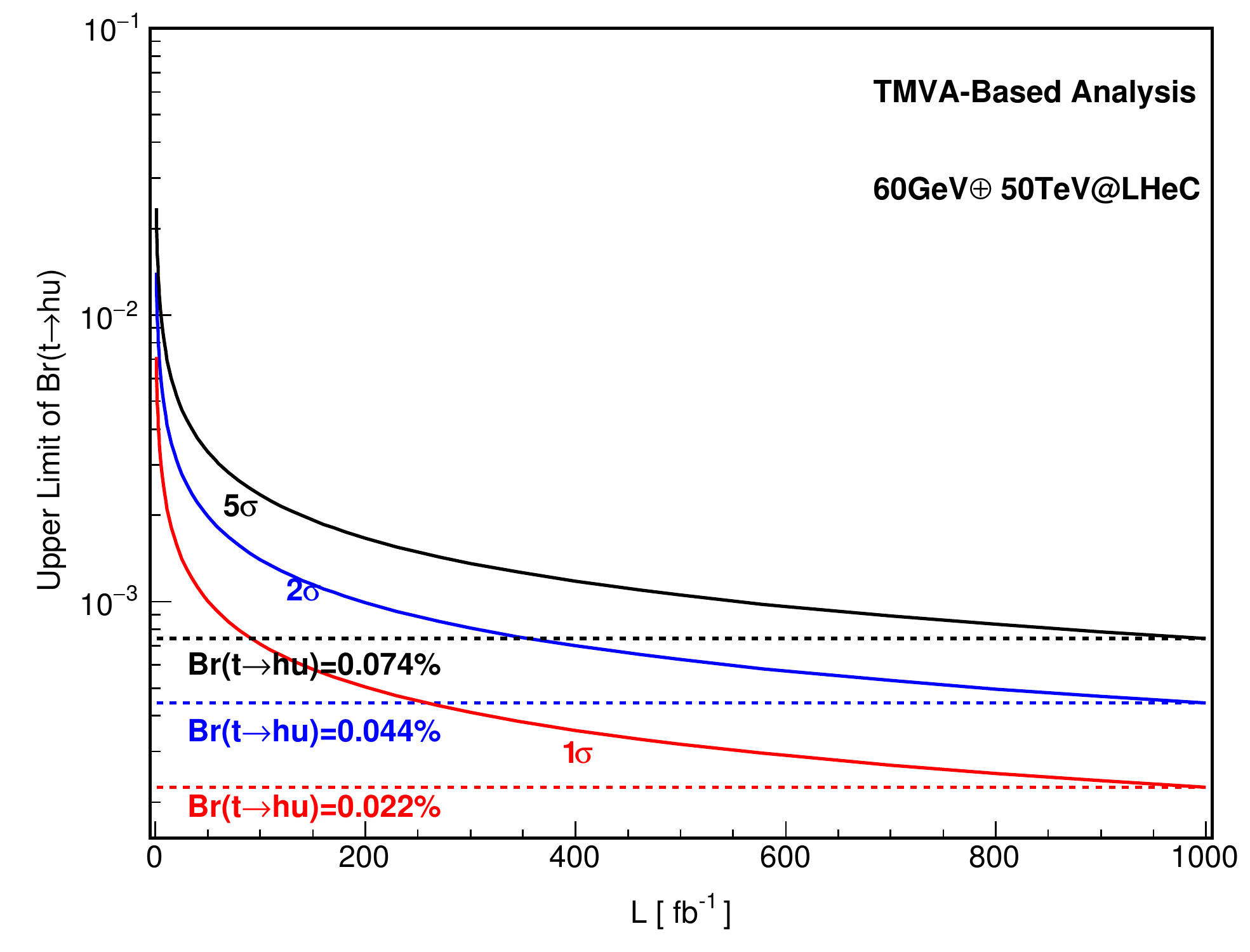}
\vspace{-0.3cm}
\caption{\label{limit}
The upper limit on $\rm Br(t\to uH)$ at 99.9, 95, 68$\%$ C.L.
as a function of the integrated luminosity at the 7(50) TeV LHeC(FCC-eh) with 60 GeV electron beam.
The red, blue and black curves present 1$\sigma$, 2$\sigma$ and 5$\sigma$ systematics.
Both limits from the Cut-based method and the TMVA based method are performed.}
\end{figure}
Our results (in Fig.\ref{limit}) show that, at the 1000 $\rm fb^{-1}$ LHeC,
the expected limit on Br($\rm t\to uH$) can be probed down to $0.113(0.093)\%$
with the Cut-based(MVA based) analysis at the $95\%$ confidence level.
The limits can be improved when polarized beam is considered\cite{tqh_pol_haosun}.
At the 50 TeV FCC-eh, the corresponding limit can be probed down to $0.03(0.022)\%$.
It is found that the potential to probe the tqH coupling can be very much improved compared to the ATLAS
and CMS experiments and even improve theoretical sensitivity of high luminosity (HL)-LHC.

Typically, the CP-nature of the ttH coupling
can also be studied through top-Higgs associated production
at the ep collider ($\rm e^-p\to \bar t H\nu_e$)\cite{TopHiggs_Kumar} (Fig.1(8)).
The CP-phase dependent lagrangian can be written here,
\begin{eqnarray}
{\cal L}  =\rm - i \frac{m_t}{v} \bar t~[\kappa \cos\zeta_t+i\gamma_5\sin\zeta_t ]t\,H.
\end{eqnarray}
Here $\rm \zeta_t$ is the phases of the top-Higgs couplings.
The case $\kappa = 1$, $\rm \zeta_t = 0$ corresponds to the SM.
$\rm \zeta_{t} = 0$ or $\rm \zeta_{t} = \pi$ correspond to a pure scalar state
while $\rm \zeta_{t} =\pi/2$ to a pure pseudo scalar state.
The ranges $\rm 0<\zeta_{t}<\pi/2$ or $\rm \pi/2<\zeta_{t}<\pi$ represent a mixture of the different CP-states.
The study was performed by considering H to $\rm b\bar{b}$ and top quark leptonic decay modes.
CP-phase dependent features are demonstrated by considering
observables like cross sections, top-quark polarisation, rapidity
difference between H and $\rm \bar{t}$ and different angular asymmetries.
\begin{figure}[hbtp]
\centering
\includegraphics[scale=0.38]{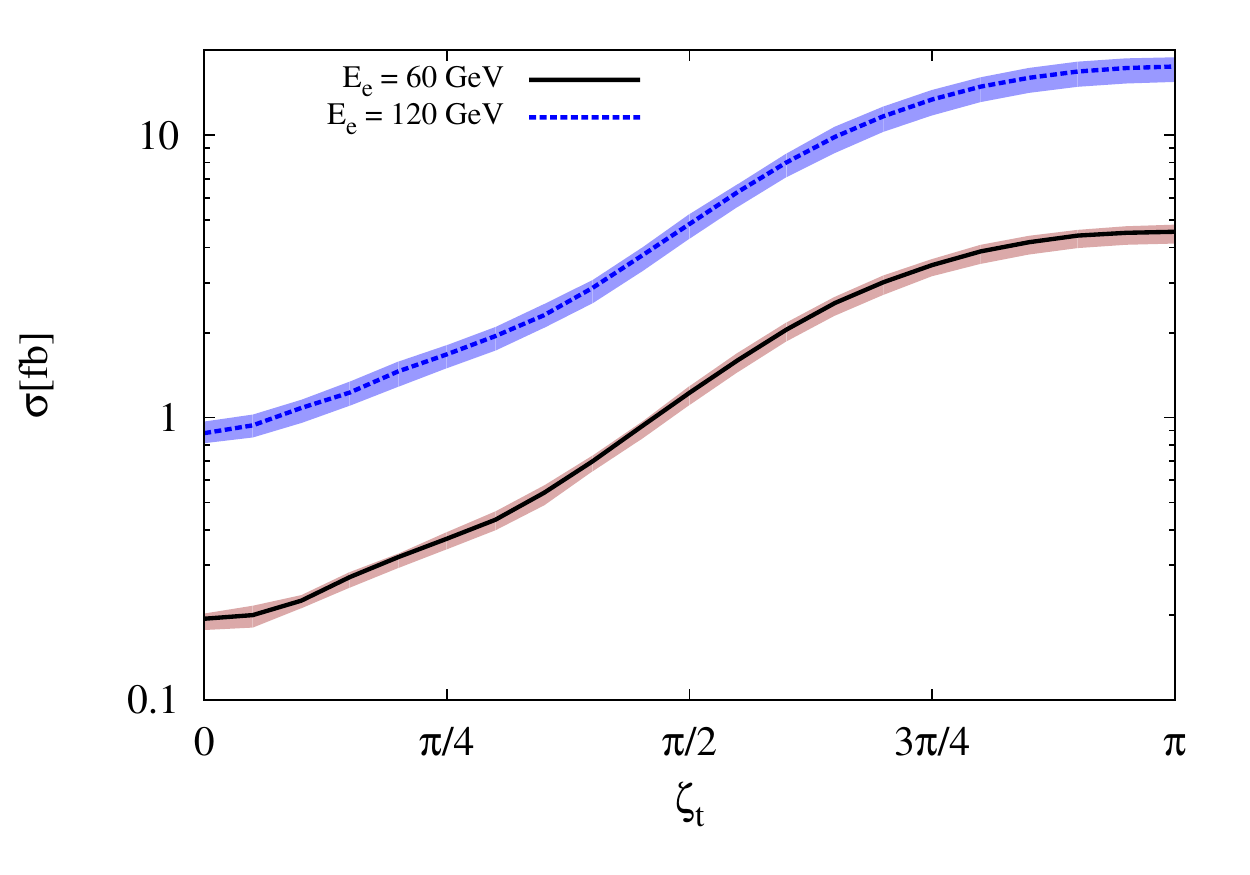}
\includegraphics[scale=0.38]{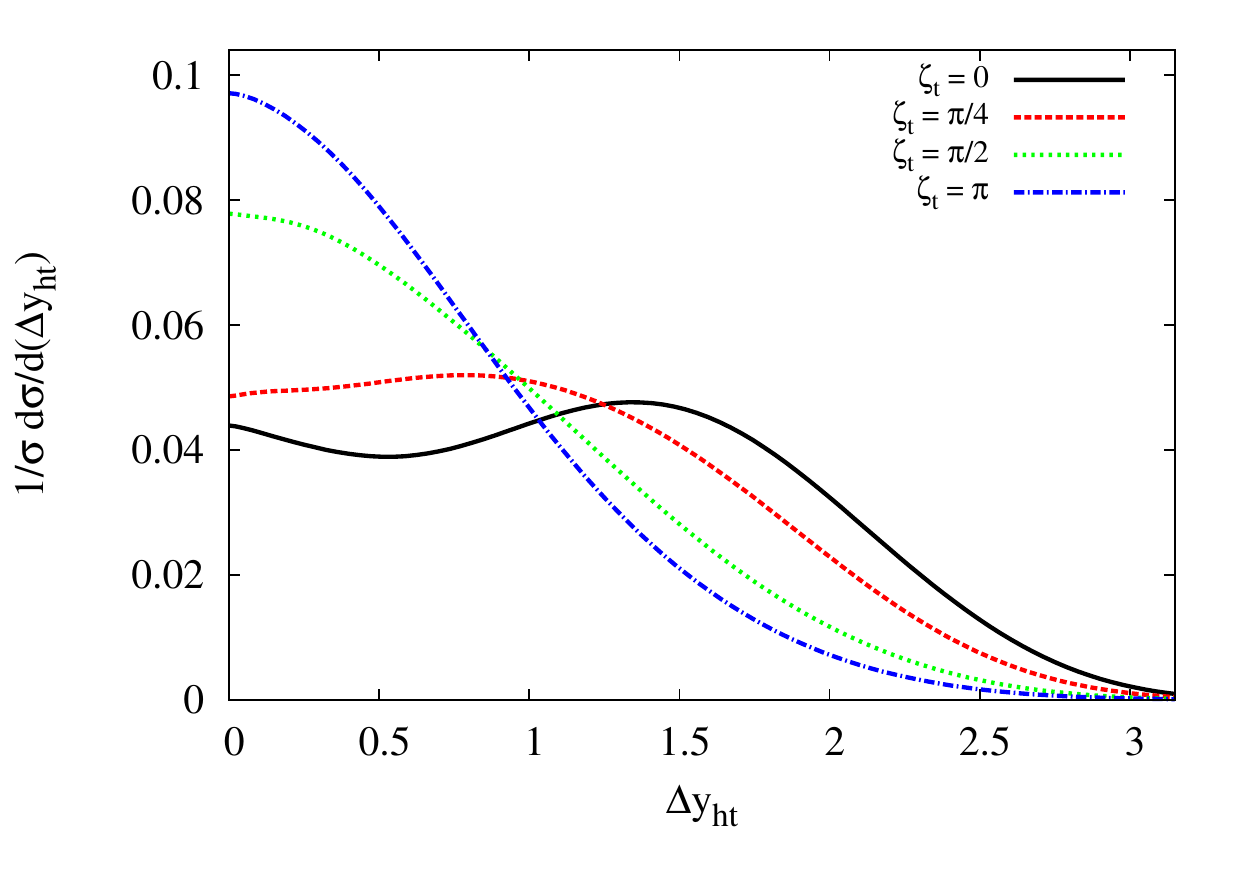}\\
\includegraphics[scale=0.38]{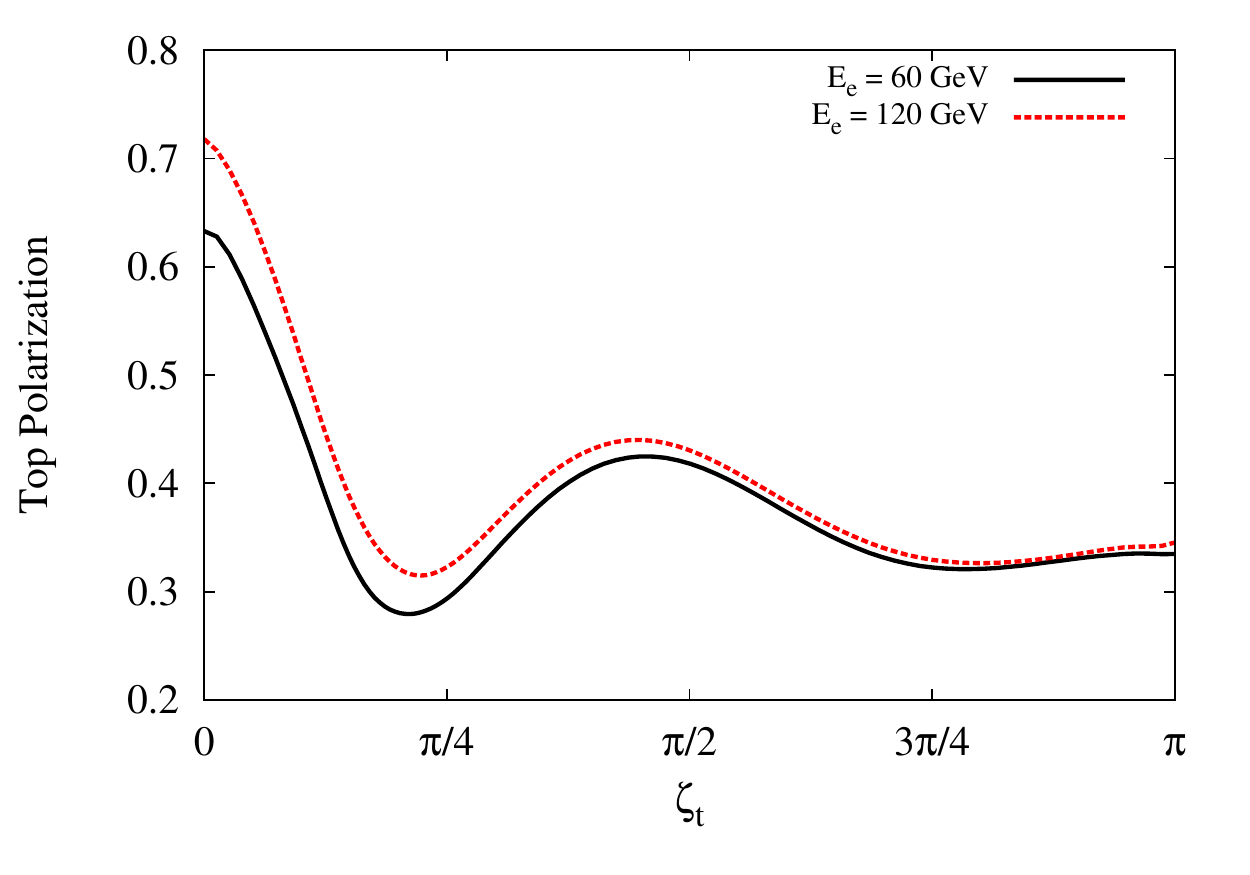}
\includegraphics[scale=0.38]{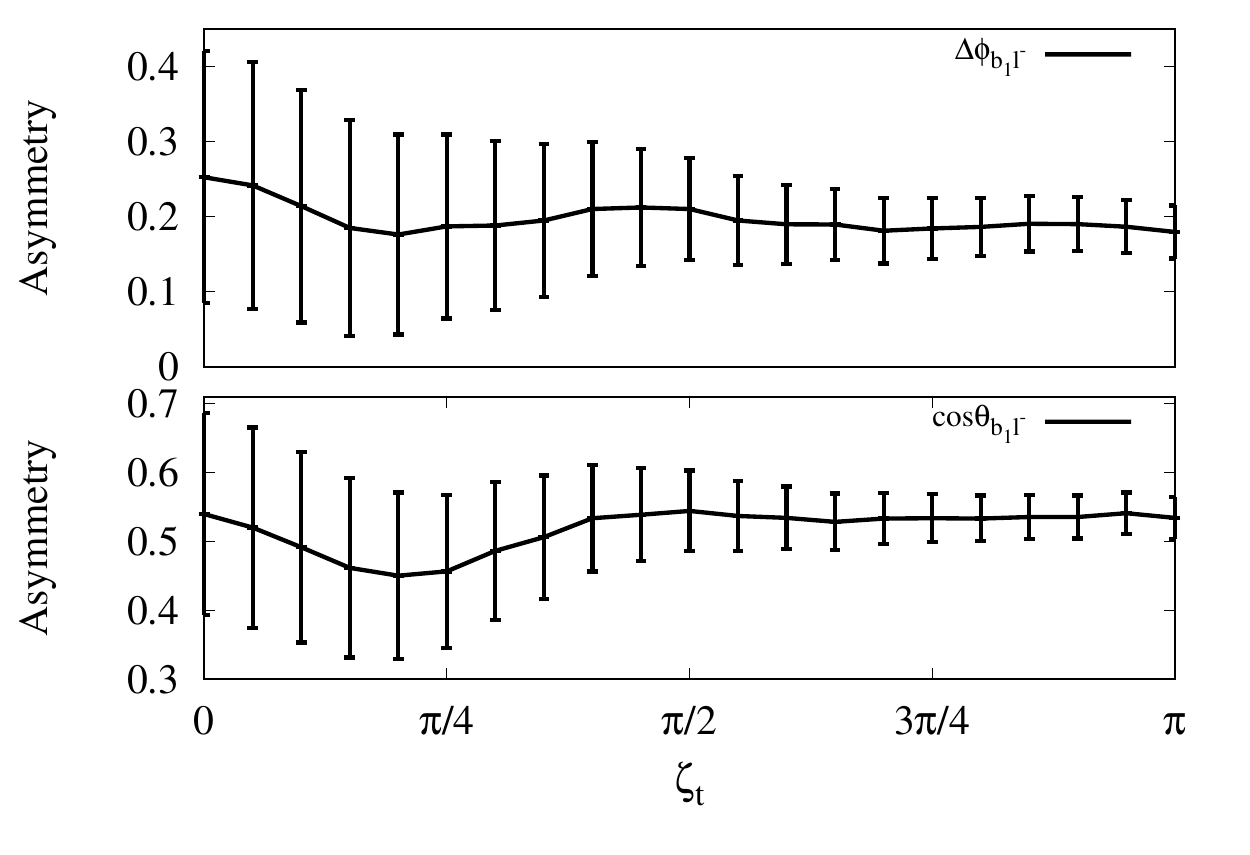}
\vspace{-0.4cm}
\caption{\label{obvervables}
Cross section as a function of $\rm \zeta_t$.
Difference between rapidities of top and Higgs at some values of $\rm \zeta_t$.
The degree of longitudinal polarisation ($\rm P_t$) of top against $\rm \zeta_t$.
Angular asymmetries between the leading b-jet and the charged lepton
in the differential azimuthal and polar angle ($\rm \Delta\phi_{b_1\ell^-}$ and $\rm \cos\theta_{b_1\ell^-}$) distributions.
}
\end{figure}
For electron and proton beam-energies of 60 GeV and 7 TeV respectively,
at luminosity $\rm L=100\ fb^{-1}$, the regions above $\rm \pi/5 < \zeta_t \leq \pi$ are
excluded at 2$\sigma$ confidence level. The accuracy of SM top-Higgs
coupling could be measured to be $\kappa = 1.00 \pm 0.17 (0.08)$
at $\rm \sqrt{s} = 1.3 (1.8)$ TeV for an ultimate $\rm L = 1\ ab^{-1}$.
Notice asymmetry studies at the HL-LHC probe $\rm \zeta_t$ up to $\pi/6$
for a total integrated luminosity of 3 $\rm ab^{-1}$.
Thus we conclude that the LHeC provides a
better environment to test the CP nature of Higgs boson couplings compare to LHC.

\section{Summary}

In this talk we present an overview of top physics at the ep colliders.
Selected topics include but not limited to top structure function, top parton distribution functions,
top spin polarization, top electric charge, measurement of $\rm V_{tb}$,
anomalous $\rm tt\gamma$, ttZ, tbW, $\rm tq\gamma$, tqH couplings and CP phase of ttH coupling, etc.
Some of these topics are being thoroughly studied with updated LHeC and FCC-eh
detector simulations and some are still updating.
New ideas or contributions are also welcomed. For example,
top-related BSM physics at the ep colliders\cite{heavyTop_Liu},
which may further emphasize the strength of the LHeC (FCC-eh) when it comes to the top sector.


\begin{thebibliography}{99}

\bibitem{top_structure_Boroun}
G.R.Boroun, \emph{Geometrical scaling behavior of the top structure functions ratio at the LHeC},
\emph{Phys.Lett.B}{\bf744} (2015) 142-145; \emph{Top structure function at the LHeC},
\emph{Phys.Lett.B}{\bf741} (2015) 197-201.

\bibitem{ttrttz_Bouzas}
A.O.Bouzas and F.Larios, \emph{Probing $tt\gamma$ and ttZ couplings at the LHeC},
\emph{Phys.Rev.D} {\bf88} (2013) 094007, [{\tt arXiv:1308.5634}].

\bibitem{tqr_Cakir}
I.T.Cakir, O.Cakir and S.Sultansoy,
\emph{Anomalous Single Top Production at the LHeC Based $\gamma$p Collider},
\emph{Phys.Lett.B} {\bf685} (2010) 170-173, [{\tt arXiv:0911.4194}].

\bibitem{LHeC_JPG2012}
LHeC Study Group, \emph{A Large Hadron Electron Collider at CERN:
Report on the Physics and Design Concepts for Machine and Detector},
\emph{J.Phys.G} {\bf39} (2012) 075001, [{\tt arXiv:1206.2913}].

\bibitem{top-spin-pol}
S.Atag and B.D.Sahin, \emph{Top quark spin in ep collision},
\emph{Phys.Rev.D} {\bf 69} (2004) 034016.

\bibitem{Wtb_Kumar}
S.Dutta, A.Goyal, M.Kumar and B.Mellado, \emph{Measuring anomalous W tb couplings
at $e^-$p collider}, \emph{Eur.Phys.J.C} {\bf 75} 12 (2015) 577, [{\tt arXiv:1307.1688}].

\bibitem{LHeC_pos}
Zhiqing Zhang, \emph{Top and EW Physics at the LHeC},
in proceedings of \emph{2015 European Physical Society Conference on High Energy Physics},
\pos{PoS(EPS-HEP2015)342} (2015), [{\tt arXiv:1511.05399}].

\bibitem{tqh_haosun}
Wei Liu, Hao Sun, XiaoJuan Wang and Xuan Luo,
\emph{Probing the anomalous FCNC top-Higgs Yukawa couplings at the Large Hadron Electron Collider},
\emph{Phys.Rev.D} {\bf 92} (2015) no.7, 074015, [{\tt arXiv:1507.03264}].

\bibitem{tqh_pol_haosun}
XiaoJuan Wang, Hao Sun and Xuan Luo,
\emph{Searches for the Anomalous FCNC Top-Higgs Couplings with Polarized Electron Beam at the LHeC},
\emph{Adv.High Energy Phys} 2017 (2017) 4693213, [{\tt arXiv:1703.02691}].

\bibitem{TopHiggs_Kumar}
Baradhwaj Coleppa, Mukesh Kumar, Satendra Kumar and Bruce Mellado,
\emph{Measuring CP nature of top-Higgs couplings at the future Large Hadron electron Collider},
\emph{Phys.Lett.B} {\bf 770} (2017) 335-341, [{\tt arXiv:1702.03426}].

\bibitem{heavyTop_Liu}
Yao-Bei Liu, \emph{Search for single production of vector-like top partners
at the Large Hadron Electron Collider}, [{\tt arXiv:1704.02059}].

\end{thebibliography}
\end{document}